\documentclass[preprint,12pt]{elsarticle}
\usepackage{amsmath}
\usepackage{amssymb}
\usepackage{hyperref}
\usepackage[para]{threeparttable}
\begin{document}

\begin{frontmatter}

\title{Co-sponsorship analysis of party politics in the 20th National Assembly
of Republic of Korea}

\author{Seung Ki Baek\corref{cor1}}
\address{Department of Physics, Pukyong National University, Busan 48513, Korea}
\ead{seungki@pknu.ac.kr}

\author{Jonghoon Kim}
\address{Department of Physics, Pukyong National University, Busan 48513, Korea}

\author{Song Sub Lee}
\address{Looxid Labs, Seoul 06247, Korea}

\author{Woo Seong Jo}
\address{
Northwestern Institute on Complex Systems, Evanston, IL 60208\\
Kellogg School of Management, Northwestern University, Evanston, IL 60208
}

\author{Beom Jun Kim\corref{cor2}}
\address{Department of Physics, Sungkyunkwan University, Suwon 16419, Korea}
\ead{beomjun@skku.edu}

\begin{abstract}
We investigate co-sponsorship among lawmakers by applying the
principal-component analysis to the bills introduced in the 20th National
Assembly of Korea. The most relevant factor for co-sponsorship is their party
membership, and we clearly observe a signal of a third-party system in action.
To identify other factors than the party influence,
we analyze how lawmakers are clustered inside each party, and the result
shows significant similarity between their committee membership and
co-sponsorship in case of the ruling party.
In addition, by monitoring each lawmaker's
similarity to the average behavior of his or her party, we have found that it
begins to decrease approximately one month before the lawmaker actually changes
the party membership.
\end{abstract}

\begin{keyword}
Legislation \sep Co-sponsorship \sep Principal-component analysis \sep
Clustering similarity
\end{keyword}

\end{frontmatter}

\section{Introduction}

Legislation is a process of cooperation and conflict: To make a law, a bill must
be moved to the Assembly by a sufficient number of lawmakers (or in some other
ways, e.g., by the government), and the bill will be passed only when approved
by a sufficient number of lawmakers. At the same time, lawmakers are divided
into parties according to their different views on public interest. Due to its
fundamental importance in democracy, this process has been recorded in a
publicly accessible form, which makes the data suitable for statistical analysis
of human interaction. For example, recent data-driven studies have revealed the
role of connectedness in the legislative
influence~\citep{fowler2006connecting,harward2010calculus,rombach2014core} and
suggested various factors playing behind the
scene~\citep{aleman2013explaining,desmarais2015measuring,montgomery2017effects,holman2018stop,aref2020detecting}
as well as quantitative measures of political
polarization~\citep{porter2007community,zhang2008community}.

In this work, we wish to provide quantitative understanding of party politics
by analyzing co-sponsorship of bills moved to the 20th National Assembly of
Republic of Korea (May 30 2016---May 29 2020). Although not many bills
become laws, such bill data express the sponsors' ideas most clearly
because bills are listed as they are introduced, compared to later stages where
the bills are revised and merged to reach a compromise. It has been reported
that party membership had the greatest influence on co-sponsorship
in the 17th Assembly~\citep{park2017cosponsorship}, and
one may ask how significant it is now, or whether other factors also explain
co-sponsorship when viewed from an ``orthogonal'' point of view. Another
related question is if party membership change is a cause or result of the
corresponding change in co-sponsorship, which we will attempt to answer after
reducing it to a simpler question, asking which one takes precedence.

Our first finding is that the 20th Assembly is best described as a third-party
system, i.e., consisting of three wings that represent liberalism, conservatism,
and centrism, respectively. Our analysis also confirms that the party influence
on co-sponsorship is absolutely dominant, and that political regionalism has
coupled lawmakers' parties to their constituencies and birthplaces. However, if
we look at the Democratic Party, which has been the ruling party since May
2017, the most relevant factor in co-sponsorship among the Democrats turns out
to be their standing committees. In addition, we have found that a lawmaker
deviates from the average direction of his or her party before leaving it, with
a time scale of one month.

\section{Datasets}

\subsection{Bill data}
The Assembly started with 300 lawmakers, and the number
has been fluctuating between 288 and 300 due to disfellowships and by-elections.
If we count all the persons who have ever been a part of the 20th Assembly
during the period, the number is 318.
We have collected their bills from People's
Solidarity for Participatory Democracy (\url{http://watch.peoplepower21.org}),
whose original source is Open Data Portal (\url{http://www.data.go.kr}).
For each bill, we have its date of motion, title, chief
author, current status, and co-sponsors' names and parties.
Although the 20th session of the Assembly was seated until 29 May 2020, we had
started analysis in the middle of the session, so
our dataset ends on June 3 2019, containing 20967 bills. The actual number used
for analysis is about 10\% less than that because we have to exclude the bills
by the government.

\subsection{Potential factors behind co-sponsorship}
We have also collected various items on individual lawmakers on the Web,
including their party membership records (\url{http://ko.wikipedia.org}), as
well as their constituencies, birthplaces, standing committees, universities,
numbers of elected terms, genders, ages, and assets
(\url{http://raythep.mk.co.kr}).
When we have to assign a unique party to each
lawmaker, we will use the one at the time of election. Independent lawmakers are
classified as a separate cluster. As a consequence, we have $K=6$ as the number
of clusters based on party membership.
As for standing committees, if a lawmaker
belongs to more than one committees, we choose the first one in the
database.
The Speaker of the National Assembly is counted as a separate committee
because he does not belong to any. As a result, the committee information
classifies lawmakers into $K=18$ clusters.
In case of birthplaces, our analysis uses the provincial-level divisions before
the foundation of Sejong Special Self-Governing City in 2007, and the resulting
number of clusters is $K=16$. The same
applies to constituencies, but this time we have to include Sejong, and
proportional representatives, who were 47 lawmakers at the
time of election, are also classified into a separate constituency. Therefore,
the number of clusters according to constituencies amounts to $K=18$ instead of
$16$.
To capture their academic ties, we have used
which universities lawmakers graduated from, with $K=66$,
because high schools are too diverse for clustering analysis. Almost
everyone has a bachelor's degree, but for those who lack it, we use their high
schools instead. The greatest number of elected terms is eight ($K=8$), and
gender has $K=2$. For ages or assets, $K$ is not determined {\it a priori} but
should be given as an input parameter. All these items are summarized in
Table~\ref{table:collection}.

\begin{table}
\caption{
List of items collected for clustering analysis on lawmakers.
For each item, we show the corresponding number of clusters, denoted by $K$.
}
      \begin{tabular}{rcl}
        \hline
        item & $K$ & misc. \\ \hline
        party & $6$ & five parties + independent lawmakers, at the time of
        election \\
        constituency & $18$ & provincial-level divisions +
        proportional representatives\\
        birthplace & $16$ & provincial-level divisions before the foundation of
        Sejong\\
        committee & $18$ & 17 standing committees + the Speaker of the National
        Assembly\\
        no. of terms & $8$ & \\
        university & $66$ & If unavailable, high-school information is used
        instead.\\
        gender & $2$ & \\
        age & N/A & $K$ is an input parameter.\\
        asset & N/A & $K$ is an input parameter. \\\hline
      \end{tabular}
      \label{table:collection}
\end{table}

\section{Methods}

In this section, we explain how we analyze the above data. Most of the
calculations have been done in the python
environment~\citep{oliphant2006guide,hunter2007matplotlib,mckinney2010data,van2011numpy}.
We have also used packages for
visualization~\citep{michael_waskom_2017_883859,plotly} and clustering
analysis~\citep{gates2019clusim}.

\subsection{Principal-component analysis (PCA)}
Assume that we have $N$ data points, each of which is $M$-dimensional. The $n$th
data point can be denoted by a column vector, $[R_{1n}, \ldots,
R_{Mn}]^\intercal$, where $\intercal$ means transpose, and the whole data can be
represented by an $M \times N$ matrix $R$.
In our case, the original data matrix $R$ has $M \approx 2 \times 10^4$
rows and $N \approx 300$ columns. Its element is binary, i.e., $R_{mn}=1$ if the
$n$th lawmaker sponsored the $m$th bill and $R_{mn}=0$
otherwise. After removing all-zero rows and columns,
we standardize the data by using the sample mean $\mu_n \equiv M^{-1}
\sum_m R_{mn}$ and standard deviation $s_n = \sqrt{\sum_m
(R_{mn}-\mu_n)^2/(M-1)}$ along the $n$th column vector. As a result, we work
with the following data matrix:
\begin{equation}
X = \begin{pmatrix}
\frac{R_{11} - \mu_1}{s_1} & \ldots & \frac{R_{1N} - \mu_N}{s_N}\\
\vdots  & \ddots & \vdots\\
\frac{R_{M1}-\mu_1}{s_1} & \ldots & \frac{R_{MN}-\mu_N}{s_N}
\end{pmatrix}.
\end{equation}
We then construct an $N \times N$ symmetric matrix $Q$ as follows:
\begin{equation}
Q = \frac{1}{M-1} X^\intercal X
= \frac{1}{M-1}
\begin{pmatrix}
X_{11} & \ldots & X_{M1}\\
\vdots & \ddots & \vdots\\
X_{1N} & \vdots & X_{MN}
\end{pmatrix}
\begin{pmatrix}
X_{11} & \ldots & X_{1N}\\
\vdots & \ddots & \vdots\\
X_{M1} & \vdots & X_{MN}
\end{pmatrix},
\end{equation}
which is called a correlation matrix. Its element $Q_{ij}$ means correlation
between the $i$th and $j$th data points:
\begin{equation}
Q_{ij} = \frac{1}{M-1} \sum_{m=1}^M X_{mi} X_{mj}
= \frac{1}{M-1} \sum_{m=1}^M \left( \frac{R_{mi}-\mu_i}{s_i} \right) \left(
\frac{R_{mj}-\mu_j}{s_j} \right),
\end{equation}
which takes a value from $[-1:1]$ with $Q_{ii} = 1$.
The key step of PCA is to diagonalize this correlation matrix $Q$, whereby we
get its eigenvalues in descending order together with the
corresponding eigenvectors. The eigenvectors are called principal axes, and
a reduced representation of the original data is obtained by taking the first
few principal axes and projecting the data onto the resulting subspace. Note
that the total sum of the eigenvalues equals $N$ in this standardized PCA.
Let us denote the $k$th eigenvalue as $\lambda_k$ and the corresponding
eigenvector as $\mathbf{e}_k$. The eigenvectors are normalized so that they form
an orthonormal set with $\mathbf{e}_k \cdot \mathbf{e}_l = \delta_{kl}$.
The correlation matrix can be decomposed in the following
way~\citep{kim2005systematic}:
\begin{equation}
Q = \sum_{k=1}^N \lambda_k \mathbf{e}_k \otimes \mathbf{e}_k,
\label{eq:decompose}
\end{equation}
where $\otimes$ means the outer product.

In performing PCA, it is usual practice to apply the singular-value
decomposition as follows:
\begin{equation}
X = U D V^\intercal,
\end{equation}
where $U$ is an $M \times M$ orthogonal matrix, $D$ is an $M \times N$
rectangular diagonal matrix with non-negative real numbers on the diagonal,
which are called singular values, and $V$ is an $N \times N$ orthogonal
matrix. The columns of $U$ are eigenvectors of $X X^\intercal$, and the
columns of $V$ are eigenvectors of $X^\intercal X$. The singular values in $D$
are the square roots of the eigenvalues of $X^\intercal X$ or $X X^\intercal$,
and the number of singular values is equal to the rank of $X$. We have the
following identities:
\begin{eqnarray}
X \mathbf{v}_k &=& \sigma_k \mathbf{u}_k\\
X^\intercal \mathbf{u}_k &=& \sigma_k \mathbf{v}_k,
\label{eq:svd_identity}
\end{eqnarray}
where $\mathbf{u}_k$ is the
$k$th column of $U$, $\sigma_k$ is the $k$th singular value, and $\mathbf{v}_k =
\mathbf{e}_k$ is the $k$th column of $V$.
Due to Eq.~\eqref{eq:svd_identity}, the $i$th data point is projected
onto the $k$th principal axis at position
\begin{equation}
z_{i,k} \equiv \sigma_k e_{i,k}.
\label{eq:zik}
\end{equation}
The distance between data points $i$ and $j$
on this axis can thus be defined as
\begin{equation}
d_{ij,k} \equiv |z_{i,k} - z_{j,k}| = \sigma_k |e_{i,k} - e_{j,k}|.
\label{eq:dijk}
\end{equation}
From Eq.~\eqref{eq:decompose}, it is straightforward to derive the following
identity:
\begin{equation}
d_{ij}^2 \equiv \sum_k d_{ij,k}^2 = 2 (1-Q_{ij}),
\label{eq:distance}
\end{equation}
where $d_{ij}$ is called correlation distance between $i$ and $j$.

\subsection{Similarity measures}
We examine the collected items one by one in the following way:
First, we assume that each given item is a ``true'' index for classification.
We then compare the result with an agglomerative clustering based on the
correlation distance [Eq.~\eqref{eq:distance}].
Ward's method is used for hierarchical linkage throughout this work.
For example, in case of gender, $N$ lawmakers form $K=2$ clusters, one for males
and the other for females. Let $c_1$ denote this ``true''
classification. We then perform agglomerative clustering until we end up with
two clusters, and denote this clustering as $c_2$.
The question is how much $c_1$ coincides with $c_2$.
It can be answered with various measures such as the Rand index, the purity
index, and normalized mutual information
(NMI)~\citep{manning2008introduction} (see also
\cite{wu2009adapting,romano2016adjusting} for comparative analyses of
clustering similarity measures),
\begin{itemize}
\item
The Rand index measures the fraction of agreements between $c_1$ and $c_2$. That
is, we begin by counting $n_+$, the number of pairs of data points that belong
to the same cluster both in $c_1$ and $c_2$, and $n_-$, the number of pairs that
belong to different clusters both in $c_1$ and $c_2$. To get the Rand index, we
divide $n_+ + n_-$ by the total number of pairs, $N(N-1)/2$.
\item
The purity index is based on the ``true'' classification $c_1$, and each
cluster in $c_2$ is given a label according to the class that is the most
frequently observed in the cluster. We count the number of data points whose
classes match with their cluster labels. The purity index is obtained by
dividing this count by $N$, the total number of data points.
\item
Mutual information (MI) measures how much information we obtain about the
classes in $c_1$ by knowing the clusters in $c_2$. We normalize MI by the
arithmetic mean between entropies of $c_1$ and $c_2$ to penalize subdividing
clusters further into smaller ones.
\end{itemize}
The statistical significance is estimated with reference to
a null model, which is generated by randomly shuffling the values of the item:
Specifically, we calculate the $p$-values by counting how many of such random
samples yield higher values than the real data. If the $p$-value is small,
observed similarity is unlikely to be an outcome of random chance. Throughout
this work, the criterion is $p \le 0.05$ and its Bonferroni-corrected versions,
and we have generated more than $10^4$ random samples to estimate each
$p$-value.
This procedure can be used for measuring similarity between
two items as well. For example, if we compare constituencies and birthplaces,
we construct two clusterings, one based on constituencies ($c_1$) and
the other based on birthplaces ($c_2$), and compute the similarity measures
between $c_1$ and $c_2$. To assess its significance, the $p$-value can be
measured, e.g., by randomly shuffling lawmakers' birthplaces.
When we compare two one-dimensional number arrays, such as comparing
between age and $z_{i,k}$ with $k$ fixed [Eq.~\eqref{eq:zik}],
we may also use Spearman's
rank correlation coefficient to avoid introducing $K$ as a free parameter.
This measure ranges from $-1$ to $1$, so we
have to check its absolute value to compute the $p$-value.

\section{Results}

  \begin{figure}
  {\centering
  \includegraphics[width=\textwidth]{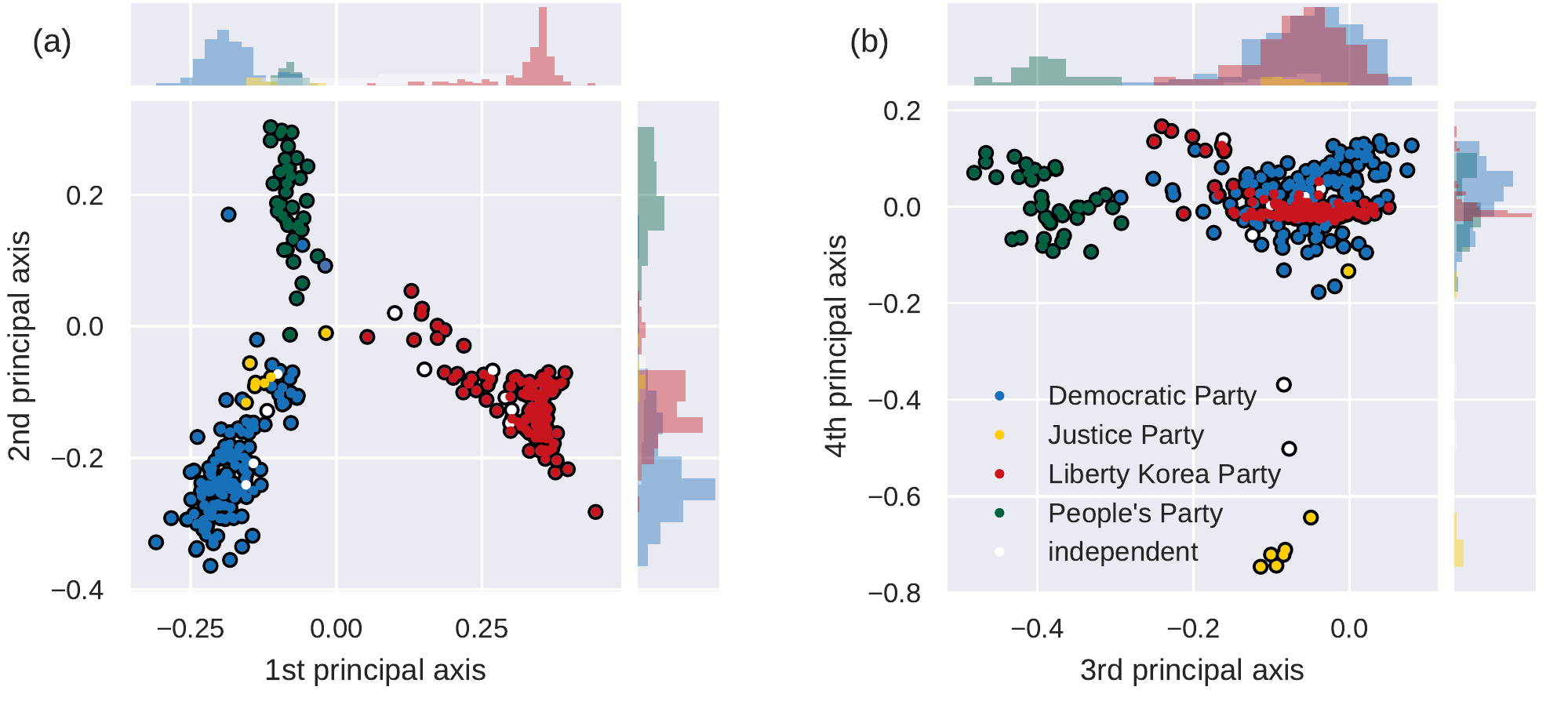}
  \par}
  \caption{Principal-component analysis of the bill data.
      (a) A tripolar structure exists when the data points are projected
      onto a two-dimensional plane spanned by the first two principal axes.
      (b) The Justice Party further separates from the Democratic Party on
      the fourth principal axis.
      }
   \label{fig:pca}
      \end{figure}

In Fig.~\ref{fig:pca}(a), we depict a two-dimensional (2D) representation of the
bill data by projecting the $N$ data points onto two principal
axes.
In this representation, they are clearly divided into three wings: The first
wing consists of two liberal parties, i.e., the Democratic Party and the
Justice Party. The second wing includes conservative ones, i.e., the Liberty
Korea Party and the Bareun Party. Finally, the third wing is comprised of the
People's Party, which claims to support centrism. Although this 2D
representation explains only 10\% of the total variance,
the result agrees with common understanding about the Assembly.
If we go further to the third and fourth principal axes,
which add about 4\% of the total variance,
the Justice Party separates from the Democratic Party
[Fig.~\ref{fig:pca}(b)].
The noticeable segregation in Fig.~\ref{fig:pca} implies that party
membership is an important predictor of co-sponsorship among lawmakers.
From this observation, we can ask the following questions:
What are the meanings of the principal axes? Which other
factors are acting on co-sponsorship? Finally, if party membership is
such a critical factor in co-sponsorship, what happens to co-sponsorship when a
lawmaker changes his or her party membership? Let us begin by answering the
second question.

\subsection{Factors behind cooperation}

\begin{figure}
  {\centering
  \includegraphics[width=0.6\textwidth]{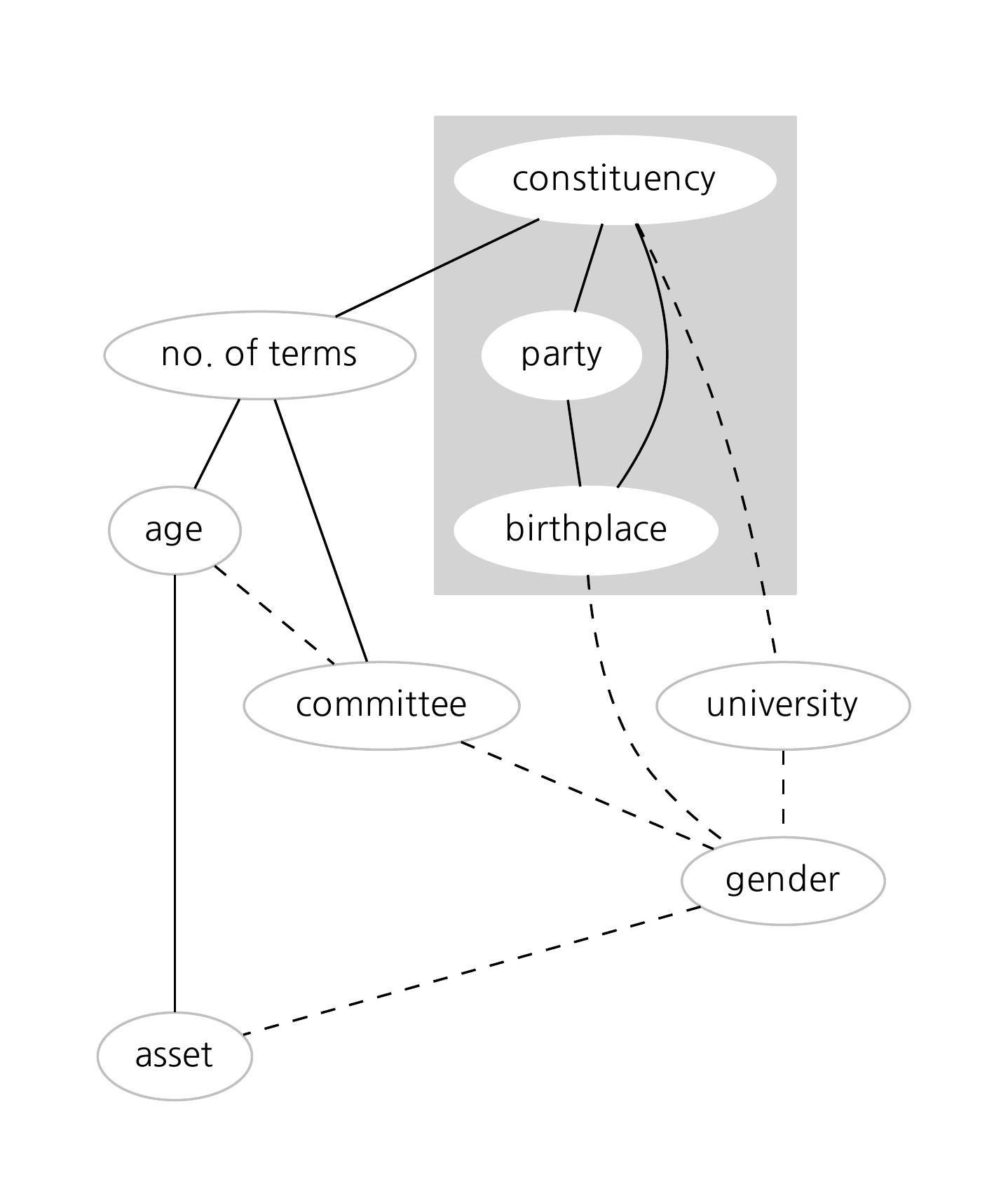}
  \par}
  \caption{Clustering similarity between each pair of items.
%      This is a graphical representation of Table~\ref{table:items}.
      Nodes are linked when their similarity is significant with respect to
      all of the following three measures: the Rand index, the purity index,
      and NMI. If two nodes are linked by a dashed line, their similarity is
      significant only when the Bonferroni correction is not applied,
      whereas a solid line means that it is significant even with the
      correction. The filled rectangle shows a clique, indicating
      strong coupling among lawmakers' parties, constituencies, and
      birthplaces.
      }
   \label{fig:items}
      \end{figure}

\begin{figure}
  {\centering
  \includegraphics[width=\textwidth]{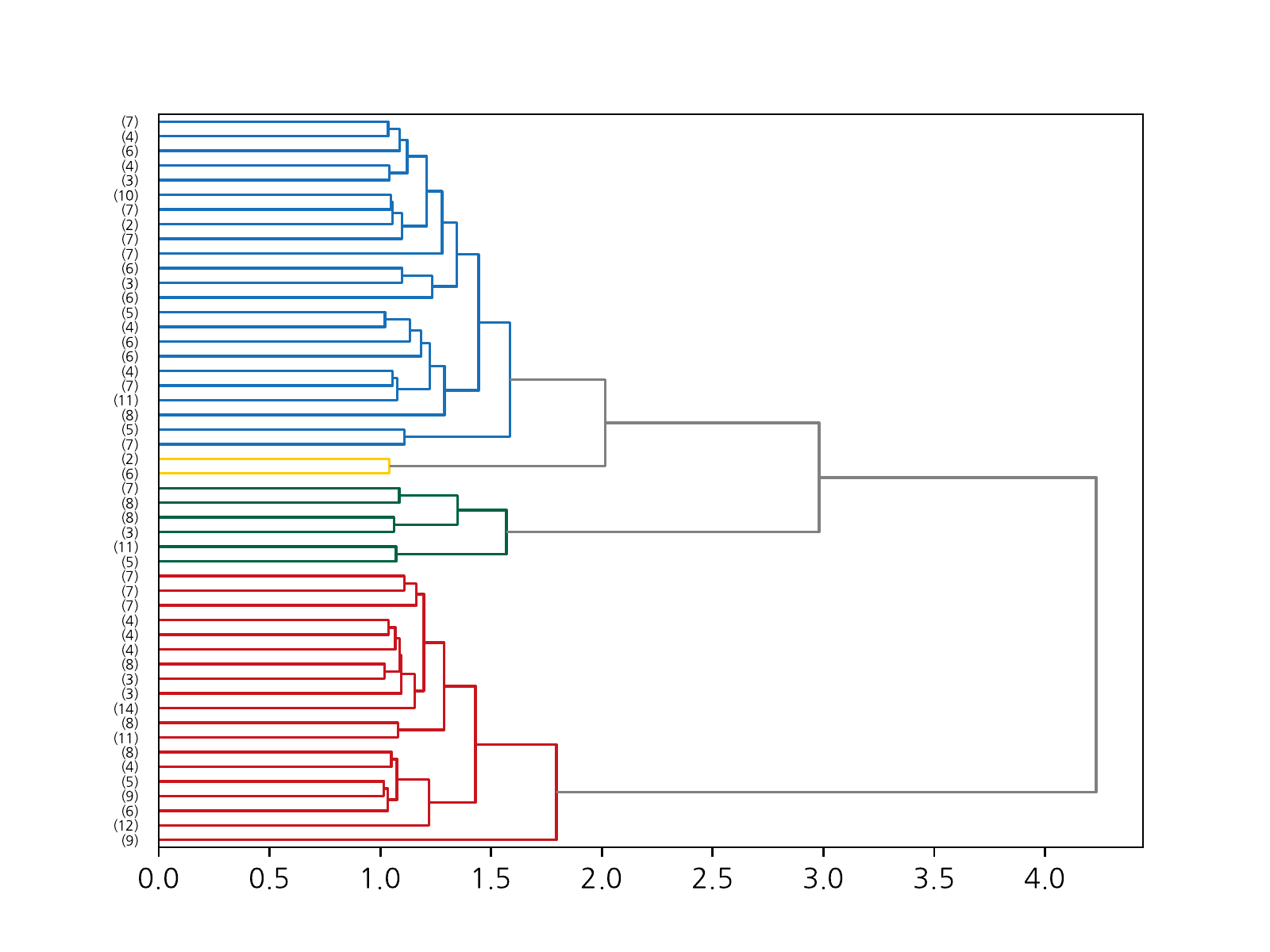}
  \par}
  \caption{Agglomerative clustering based on correlation distance.
      Each number on the vertical axis means the number of
      lawmakers belonging to the corresponding branch, and these lawmakers'
      party membership is represented by the color of the branch as in
      Fig.~\ref{fig:pca}. We use Ward's method for hierarchical
      linkage throughout this work.
      }
   \label{fig:dendrogram}
      \end{figure}

To begin with, we check similarity between each pair of items, that is,
how one item is similar to another when used as indices for clustering
lawmakers. As explained in the previous section, we will use three well-known
similarity measures: the Rand index, the purity index, and NMI.
As shown in
%Table~\ref{table:items} and
Fig.~\ref{fig:items}, lawmakers' parties,
constituencies, and birthplaces are strongly coupled in terms of all these three
measures, and this coupling can be attributed to political regionalism.

\begin{table}
\caption{
Similarity between item-based classification and correlation-based clustering
(Fig.~\ref{fig:dendrogram}).
The measures are calculated for the whole Assembly and then for three major
parties: the Democratic Party, the Liberty Korea Party, and the People's Party.
We write `Yes' if $p \le 0.05$ for the Rand index, the purity index, and
NMI, whose values are given as footnotes. The dagger symbol means that the
similarity is not judged as significant if we apply the Bonferroni correction
($p \le 0.0015$).
For ages and assets, we have attempted the agglomerative
clustering with varying the number of clusters from $K=2$ to $10$.
If similarity is significant for multiple values of $K$, we show the case
of the smallest $K$.
}
      \begin{threeparttable}
      \begin{tabular}{r|cccc}
        \hline
        item \textbackslash range & Assembly & Democratic Party & Liberty Korea
        Party & People's Party \\\hline\hline
        party        & Yes\tnote{a} & N/A & N/A & N/A \\\hline
        constituency & Yes\tnote{b} & & Yes\tnote{c} &
        $^\dagger$Yes\tnote{d}\\\hline
        birthplace   & Yes\tnote{e} & & $^\dagger$Yes\tnote{f} & \\\hline
        committee    & & $^\dagger$Yes\tnote{g} & & \\\hline
        no. of terms & & & & \\\hline
        university   & & & & \\\hline
        gender       & & & & \\\hline
        age ($K=2$)  & $^\dagger$Yes\tnote{h} & & & \\\hline
        asset        & & & & \\\hline
      \end{tabular}
      \begin{tablenotes}
      \item[a] $(0.9,1.0,0.8)$
      \item[b] $(0.8,0.3,0.3)$
      \item[c] $(0.9,0.3,0.3)$
      \item[d] $(0.7,0.5,0.3)$
      \item[e] $(0.8,0.2,0.2)$
      \item[f] $(0.8,0.3,0.3)$
      \item[g] $(0.9,0.3,0.4)$
      \item[h] $(0.5,0.6,0.01)$
      \end{tablenotes}
      \end{threeparttable}
      \label{table:corr_clustering}
\end{table}

Then, in Table~\ref{table:corr_clustering}, we compare each item-based
classification with the correlation-based agglomerative clustering
(Fig.~\ref{fig:dendrogram}), keeping the number of clusters the same on both
sides.
If we consider the whole Assembly ($N=318$), parties show significant similarity
to the agglomerative clustering with respect all the three measures. Similarity
is also significant when constituencies or birthplaces are used, which is
expected from their strong coupling with parties
%(Table~\ref{table:items} and
(Fig.~\ref{fig:items}). In addition,
we have found that clusters in the Assembly bears similarity to the age
structure (Table~\ref{table:corr_clustering}).
If we use $K=2$, for example, the $p$-value is estimated as
approximately $0.02$ for every measure.

At the same time, differences do exist among parties.
In Table~\ref{table:corr_clustering}, we have listed the cases within
three major parties, constituting the three wings of Fig.~\ref{fig:pca}(a).
Whereas co-sponsorship yields the most similar clustering to that
of constituencies in the Liberty Korea Party ($N=126$) as well as in the
People's Party ($N=38$), the lawmakers in the Democratic Party ($N=136$) show a
degree of similarity in co-sponsorship only when compared with their
committees ($p \approx 0.01$ for every measure).
Table~\ref{table:corr_clustering} also shows that the similarity
to age, observed on the Assembly level, disappears when we look at the parties,
which implies that the similarity is related with the difference between `young'
and `old' parties.

\subsection{Making sense of principal axes}
\begin{figure}
  {\centering
  \includegraphics[width=0.8\textwidth]{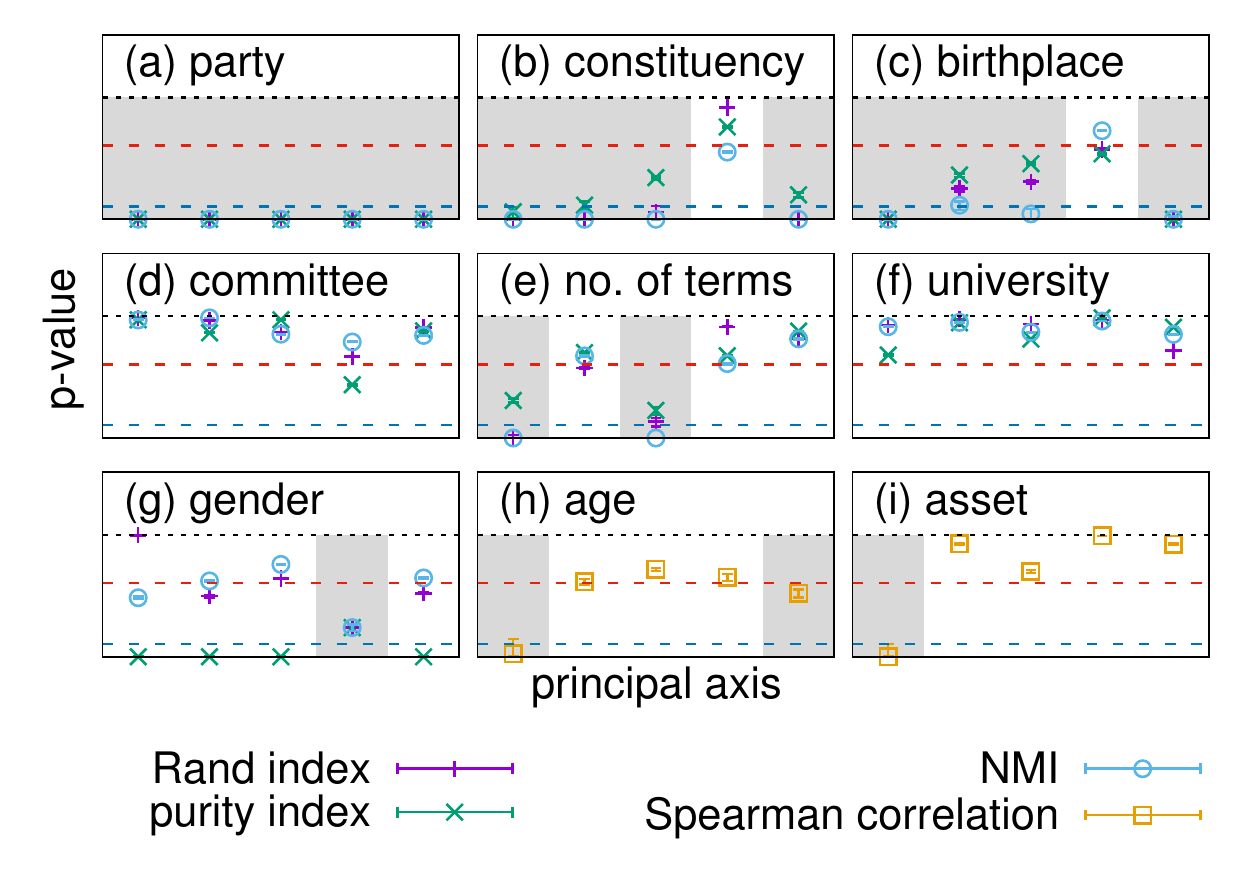}
  \par}
  \caption{Clustering similarity between each item and distance along
  each principal axis.
      We perform agglomerative clustering based on $d_{ij,k}$, the distance
      between $i$ and $j$ on the $k$th principal axis.
      Each horizontal axis shows $k$, the index for principal axes, and the
      vertical one shows $p$-values of clustering
      similarity measures in the logarithmic scale.
      From (a) to (g), we measure the Rand index, the purity index, and NMI,
      whereas Spearman's rank correlation coefficient is calculated for (h) and
      (i).
      The horizontal lines represent $p=1$, $p=0.05$, and $p=0.001$
      (Bonferroni-corrected), respectively, from top to bottom.
      We observe $p \le 0.05$ for every measure in the shaded regions.
      }
   \label{fig:axes}
      \end{figure}

So far, we have performed the agglomerative clustering based on $d_{ij}$
[Eq.~\eqref{eq:distance}]. Here, we will apply the same method to $d_{ij,k}$ in
Eq.~\eqref{eq:dijk} across different principal axes with $k=1,\ldots,5$.
For ages or assets, it is more convenient to calculate Spearman's rank
correlation coefficient with $z_{i,k}$ [Eq.~\eqref{eq:zik}].
As depicted in Fig.~\ref{fig:axes}, the party influence is pervasive along all
those principal axes.
Considering the strong coupling of parties to
constituencies and birthplaces
(see the filled rectangle in Fig.~\ref{fig:items}),
it should not be surprising that a substantial
amount of overlap exists from Fig.~\ref{fig:axes}(a) to (c).
Age is also found significant on the first principal axis
[Fig.~\ref{fig:axes}(h)], based on which `young' and `old' parties are separated
[Fig.~\ref{fig:pca}(a)].
Also by considering that a lawmaker's number of terms and assets are correlated
with his or her age (Fig.~\ref{fig:items}),
%and Table~\ref{table:items}),
we can
explain the reason that these two items are also significant on the first
principal axis [Figs.~\ref{fig:axes}(e) and (i)]. We can thus say that the first
two principal axes reflect the configuration of parties and other closely
related items. For the next two principal axes, on the other hand, we observe
significant signal in the number of terms and gender [Fig.~\ref{fig:axes}(e) and
(g)], which we have to examine more closely.

If we check how lawmakers are clustered on the third principal axis,
we find a nontrivial unimodal structure in their numbers of terms:
Among the $K=8$ clusters for this item (Table~\ref{table:collection}),
the two leftmost clusters consist almost exclusively of the
People's Party [see Fig.~\ref{fig:pca}(b) along the horizontal axis],
whose average number of terms is $T \approx 1.9$.
The other six clusters on the right are mixtures of the other parties, and
the interface between the People's Party and the rest is
occupied by lawmakers elected for many terms, so the fourth cluster in the
middle has the greatest number of terms, $T \approx 3.2$ on average.
It decreases again as we go across the interface, so the rightmost cluster, on
the opposite side of the People's Party, is composed of newly-elected lawmakers
with the smallest $T \approx 1.5$. To sum up, the third principal axis shows
statistical significance of a ``normal mode'', in which the newly-elected
lawmakers in the People's Party move in an anti-correlated manner with respect
to those who are newly elected in the other parties.

As for gender, its significance on the fourth principal axis does not imply any
possibility of non-partisan gender politics, as one can still see the party
influence there [Fig.~\ref{fig:axes}(a)]. To understand the origin of this
significance, one should note that the Justice Party separates from the others
on the fourth principal axis [Fig.~\ref{fig:pca}(b)]. The significance of gender
is due to the fact that it also has a much higher fraction of female lawmakers
($\sim 50\%$) than those of the other parties ($10 \sim 20 \%$). This effect was
negligible in clustering the whole Assembly
(Table~\ref{table:corr_clustering}) because the Justice Party has a small number
of lawmakers. Moreover, if we look closely at the agglomerative clustering on
the fourth principal axis (not shown), a female block also exists inside the
Democratic Party, but its position is almost on the opposite side of the Justice
Party.
Therefore, we conclude that gender itself is not a meaningful indicator to
interpret the position on the fourth principal axis.
We can simply say that the Justice Party is a separate entity from the
Democratic Party and others, and that its uniqueness is captured only by
gender as far as we restrict ourselves to the items of
Table~\ref{table:collection}.

\subsection{Party membership change}
  \begin{figure}
  {\centering
  \includegraphics[width=\textwidth]{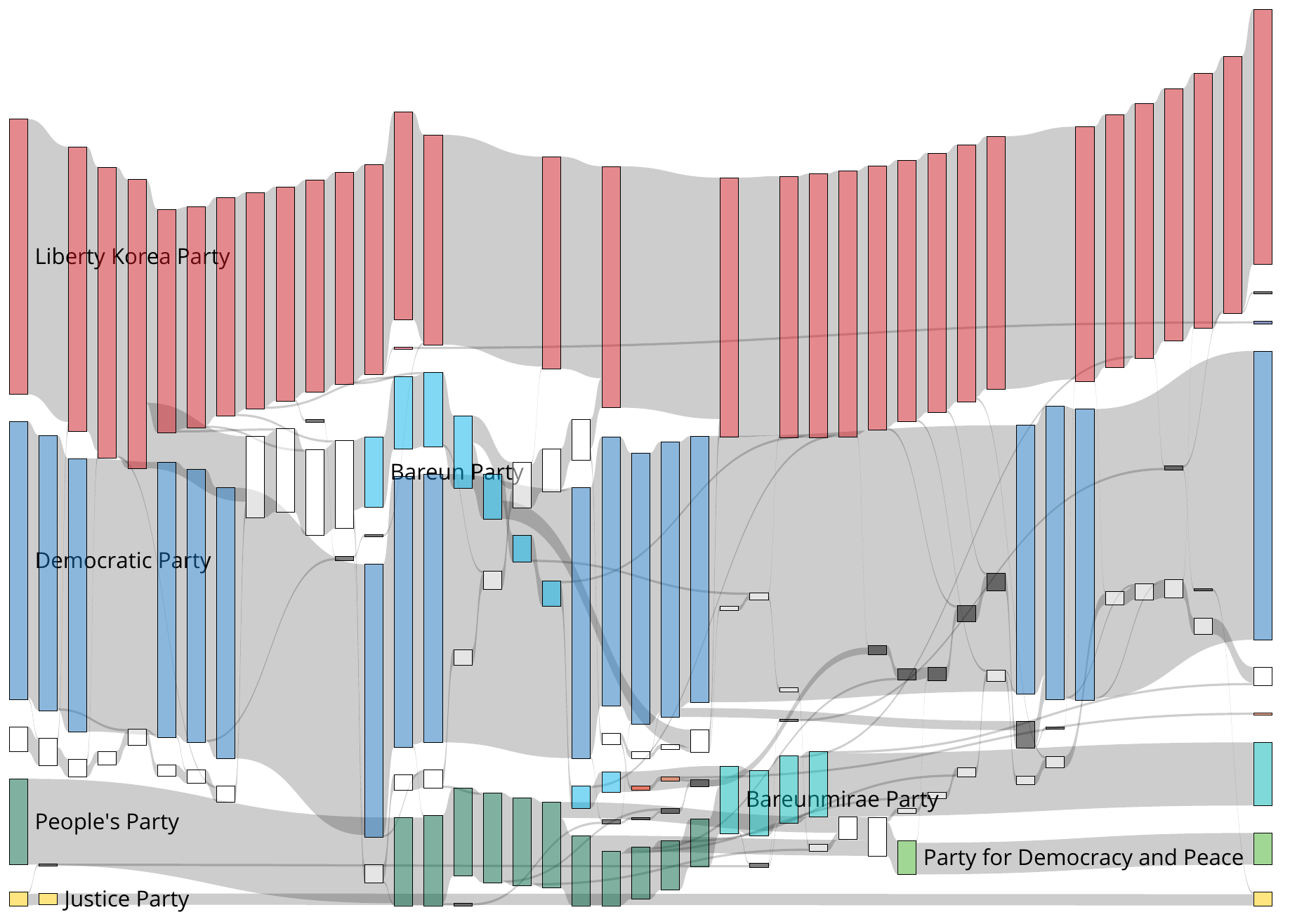}
  \par}
  \caption{Changes in the alignment of parties.
   A white node means a group of independent lawmakers with no party membership,
   and a black node means a group of vacancies to be filled by by-elections.
   Otherwise, nodes are filled with the colors of the corresponding parties.
   The centrism wing (green in Fig.~\ref{fig:pca}) has undergone substantial
   changes, among which the biggest one can be written as follows: $\text{Bareun
   Party} + \text{People's Party} \to \text{Bareunmirae Party} + \text{Party
   for Democracy and Peace}$.
   }
   \label{fig:align}
      \end{figure}

The wings in Fig.~\ref{fig:pca} have relatively stable directions throughout
the observation period. At the same time, as depicted in
Fig.~\ref{fig:align}, we observe a number of events that
lawmakers change their party membership. If a co-sponsorship network also
changes with such an event, which is entirely plausible, the question is which
one goes first.

  \begin{figure}
  {\centering
  \includegraphics[width=0.6\textwidth]{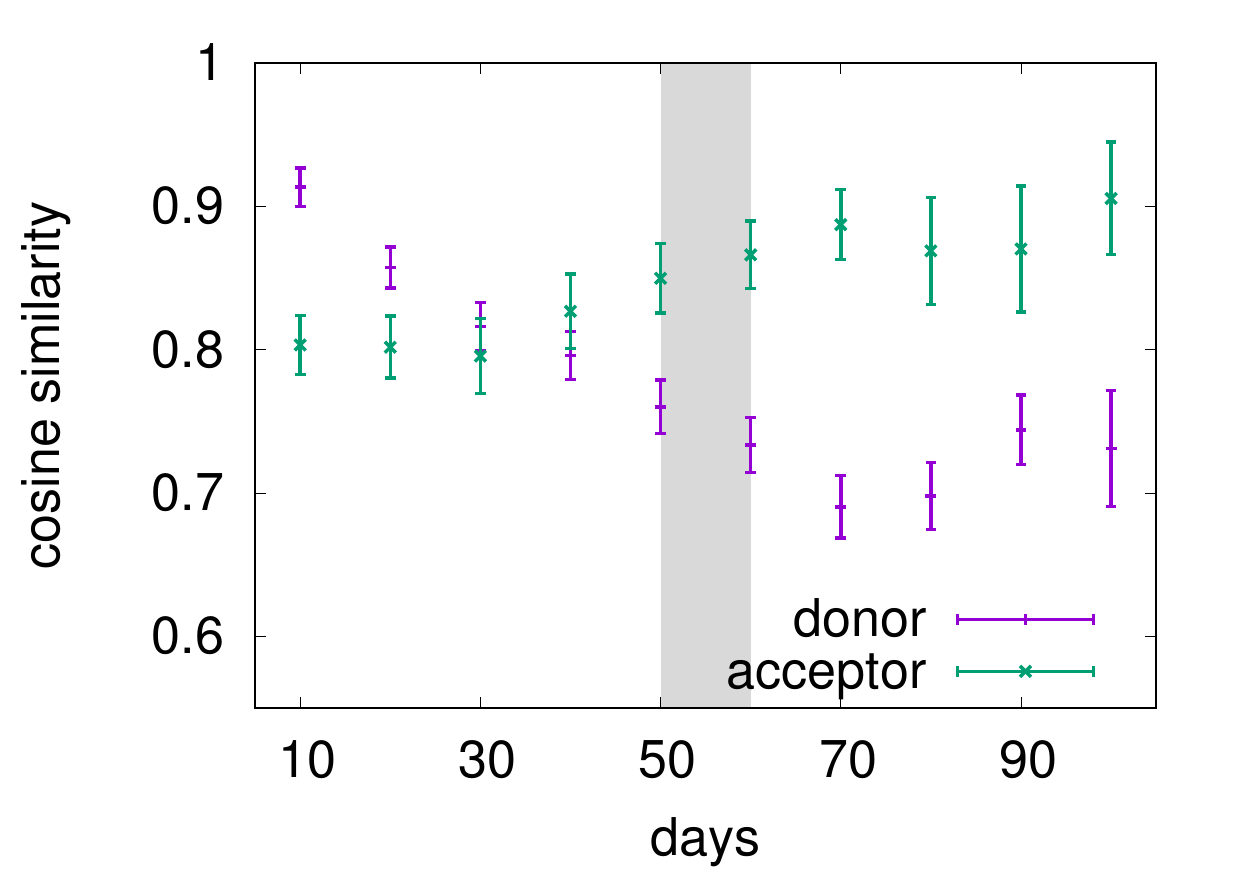}
  \par}
  \caption{Behavioral readjustment of lawmakers changing party
  membership.
      We measure lawmakers' average cosine similarity in comparison
      with their respective donor and acceptor parties, together with standard
      error. The first four principal axes have been used to calculate their
      directions (Fig.~\ref{fig:pca}).
      Each error bar contains $60 \sim 200$ points, and the shaded
      rectangle means the transition region where they change the membership.
      }
   \label{fig:change}
      \end{figure}
This question can be answered by utilizing PCA.
For each lawmaker changing the party membership, we take a time window of 100
days, within which PCA is performed to detect the following three vectors
(see Fig.~\ref{fig:pca}): The average direction of the `donor' party, denoted by
$\mathbf{d}$, that of the `acceptor' party, $\mathbf{a}$, and the lawmaker's
individual direction $\mathbf{w}$. Let us define cosine similarity between two
nonzero vectors $\mathbf{A}$ and $\mathbf{B}$ as follows:
\begin{equation}
C(\mathbf{A}, \mathbf{B}) \equiv \frac{\mathbf{A} \cdot \mathbf{B}}{|\mathbf{A}|
|\mathbf{B}|}.
\end{equation}
We measure this quantity between $\mathbf{d}$ and $\mathbf{w}$ as well as
between $\mathbf{a}$ and $\mathbf{w}$ to see how they evolve as the time window
moves. The measurement is averaged over membership-changing events. The result
in Fig.~\ref{fig:change} shows a crossing of $C(\mathbf{d}, \mathbf{w})$ and
$C(\mathbf{a}, \mathbf{w})$ about 20--30 days before the actual event.

\section{Summary and Discussion}

In summary, we have analyzed co-sponsorship among lawmakers in the 20th
National Assembly of Korea to understand how party politics works, such as how
lawmakers conform to the directions of their parties and which factors
complement the party influence.

Despite the short history and relatively small size, the centrism wing has
proved its own identity in legislation [Figs.~\ref{fig:pca}(a) and
\ref{fig:dendrogram}].
As Duverger's law states~\citep{riker1982two}, such a third-party
system had long been regarded as unstable in Korea before the 20th Assembly.
Indeed, it was for the
first time in 16 years that the Assembly started with three negotiation bodies,
and the centrism wing was even broken into two parties in the middle of our
observation period (Fig.~\ref{fig:align}). The two parties nevertheless
constituted a single wing, keeping the same direction as before, till the end of
our observation.
Even with their own political identity, the centrism wing is
losing support, and the overall tendency has indicated that it is on the process
of disappearance in agreement with Duverger's law: The size shrank further down
to about 20 lawmakers in February 2020, when another party alignment happened in
the centrism wing, and they won only three seats in the 21st legislative
elections on 15 April 2020.

Besides party membership, we have found that age yields a similar clustering to
the one that is based on co-sponsorship (Table~\ref{table:corr_clustering}),
but it seems to reflect the age difference among parties. If we look inside each
party,
the committee membership serves as a significant index in clustering the
Democrats. In addition,
our clustering analysis, combined with PCA, has detected an anti-correlated mode
of newly-elected lawmakers [Fig.~\ref{fig:axes}(e)], and this mode provides
possible interpretation for the third principal axis in Fig.~\ref{fig:pca}(b).
It is still subject to the party influence in the sense that the mode
gives a unique status to the People's Party, but the shape of the mode suggests
how the other parties have collectively responded to the emergence of the
centrism wing.
On the other hand, the fourth principal axis shows cohesion of the Justice Party
in co-sponsorship. Here, we interpret the statistical significance of gender
[Fig.~\ref{fig:axes}(g)] as meaning that the Justice Party
is uniquely characterized only by gender among our collected items.
Such a gender characteristic of the Justice Party is
related to the fact that most of its seats were awarded through proportional
representation, for which the law stipulates that the fraction of female
candidates must be no less than $50\%$.
What is absent can be another meaningful piece of
information: We do not observe any significance in academic ties
[Fig.~\ref{fig:axes}(f)].

Finally, provided that party membership is tightly bound to co-sponsorship,
we have seen that individual change of co-sponsorship comes roughly one month
before the actual membership change
(Fig.~\ref{fig:change}). Admittedly, it is hard to establish causal relationship
between the two events because one could decide inwardly to go over to another
party well before changing the co-sponsorship. Still, the change in
co-sponsorship can serve as an early indicator which tacitly but patently
manifests the decision before taking the crucial step.

Legislation exerts far-reaching impacts on the whole society, and understanding
its working mechanism is gaining more and more importance in shaping our lives.
Our analysis demonstrates that data-driven studies will help us move beyond
anecdotes and deepen our understanding of politics in a quantitative way.

\section*{Acknowledgments}
We are grateful to Woncheol Jang for his careful reading and helpful comments.
S.K.B. was supported by Pukyong National University Research Fund in 2019.
B.J.K. was supported by the National Research Foundation of Korea (NRF) grant
funded by the Korea government (MSIT) Grant~No.~2019R1A2C2089463.

%\bibliographystyle{elsarticle-num}
%\bibliography{party}

\begin{thebibliography}{10}
\expandafter\ifx\csname url\endcsname\relax
  \def\url#1{\texttt{#1}}\fi
\expandafter\ifx\csname urlprefix\endcsname\relax\def\urlprefix{URL }\fi
\expandafter\ifx\csname href\endcsname\relax
  \def\href#1#2{#2} \def\path#1{#1}\fi

\bibitem{fowler2006connecting}
J.~H. Fowler, Connecting the {C}ongress: A study of cosponsorship networks,
  Political Anal. 14~(4) (2006) 456--487.
\newblock \href {https://doi.org/https://doi.org/10.1093/pan/mpl002}
  {\path{doi:https://doi.org/10.1093/pan/mpl002}}.

\bibitem{harward2010calculus}
B.~M. Harward, K.~W. Moffett, The calculus of cosponsorship in the {US Senate},
  Legis. Stud. Q. 35~(1) (2010) 117--143.
\newblock \href {https://doi.org/https://doi.org/10.3162/036298010790821950}
  {\path{doi:https://doi.org/10.3162/036298010790821950}}.

\bibitem{rombach2014core}
M.~P. Rombach, M.~A. Porter, J.~H. Fowler, P.~J. Mucha, Core-periphery
  structure in networks, SIAM J. Appl. Math. 74~(1) (2014) 167--190.
\newblock \href {https://doi.org/https://doi.org/10.1137/120881683}
  {\path{doi:https://doi.org/10.1137/120881683}}.

\bibitem{aleman2013explaining}
E.~Alem{\'a}n, E.~Calvo, Explaining policy ties in presidential congresses: A
  network analysis of bill initiation data, Political Stud. 61~(2) (2013)
  356--377.
\newblock \href
  {https://doi.org/https://doi.org/10.1111\%2Fj.1467-9248.2012.00964.x}
  {\path{doi:https://doi.org/10.1111\%2Fj.1467-9248.2012.00964.x}}.

\bibitem{desmarais2015measuring}
B.~A. Desmarais, V.~G. Moscardelli, B.~F. Schaffner, M.~S. Kowal, Measuring
  legislative collaboration: The {Senate} press events network, Soc. Netw. 40
  (2015) 43--54.
\newblock \href {https://doi.org/https://doi.org/10.1016/j.socnet.2014.07.006}
  {\path{doi:https://doi.org/10.1016/j.socnet.2014.07.006}}.

\bibitem{montgomery2017effects}
J.~M. Montgomery, B.~Nyhan, The effects of congressional staff networks in the
  {US House of Representatives}, J. Politics 79~(3) (2017) 745--761.
\newblock \href {https://doi.org/https://doi.org/10.1086/690301}
  {\path{doi:https://doi.org/10.1086/690301}}.

\bibitem{holman2018stop}
M.~R. Holman, A.~Mahoney, Stop, collaborate, and listen: Women's collaboration
  in {US} state legislatures, Legis. Stud. Q. 43~(2) (2018) 179--206.
\newblock \href {https://doi.org/https://doi.org/10.1111/lsq.12199}
  {\path{doi:https://doi.org/10.1111/lsq.12199}}.

\bibitem{aref2020detecting}
S.~Aref, Z.~Neal, Detecting coalitions by optimally partitioning signed
  networks of political collaboration, Sci. Rep. 10~(1) (2020) 1--10.
\newblock \href {https://doi.org/https://doi.org/10.1038/s41598-020-58471-z}
  {\path{doi:https://doi.org/10.1038/s41598-020-58471-z}}.

\bibitem{porter2007community}
M.~A. Porter, P.~J. Mucha, M.~E. Newman, A.~J. Friend, Community structure in
  the {United States House of Representatives}, Physica A 386~(1) (2007)
  414--438.
\newblock \href {https://doi.org/https://doi.org/10.1016/j.physa.2007.07.039}
  {\path{doi:https://doi.org/10.1016/j.physa.2007.07.039}}.

\bibitem{zhang2008community}
Y.~Zhang, A.~J. Friend, A.~L. Traud, M.~A. Porter, J.~H. Fowler, P.~J. Mucha,
  Community structure in {Congressional} cosponsorship networks, Physica A
  387~(7) (2008) 1705--1712.
\newblock \href {https://doi.org/https://doi.org/10.1016/j.physa.2007.11.004}
  {\path{doi:https://doi.org/10.1016/j.physa.2007.11.004}}.

\bibitem{park2017cosponsorship}
C.~Park, W.~Jang, Cosponsorship networks in the 17th {National Assembly of
  Republic of Korea}, Korean J. Appl. Stat. 30~(3) (2017) 403--415.
\newblock \href {https://doi.org/https://doi.org/10.5351/KJAS.2017.30.3.403}
  {\path{doi:https://doi.org/10.5351/KJAS.2017.30.3.403}}.

\bibitem{oliphant2006guide}
T.~E. Oliphant, Guide to NumPy, MIT Press, Cambridge, MA, 2006.

\bibitem{hunter2007matplotlib}
J.~D. Hunter, Matplotlib: A {2D} graphics environment, Computing in science \&
  engineering 9~(3) (2007) 90.
\newblock \href {https://doi.org/https://doi.org/10.1109/MCSE.2007.55}
  {\path{doi:https://doi.org/10.1109/MCSE.2007.55}}.

\bibitem{mckinney2010data}
W.~McKinney, et~al., Data structures for statistical computing in python, in:
  Proceedings of the 9th Python in Science Conference, Vol. 445, Austin, TX,
  2010, pp. 51--56.

\bibitem{van2011numpy}
S.~Van Der~Walt, S.~C. Colbert, G.~Varoquaux, The numpy array: a structure for
  efficient numerical computation, Computing in Science \& Engineering 13~(2)
  (2011) 22.
\newblock \href {https://doi.org/http://dx.doi.org/10.1109/MCSE.2011.37}
  {\path{doi:http://dx.doi.org/10.1109/MCSE.2011.37}}.

\bibitem{michael_waskom_2017_883859}
M.~Waskom, O.~Botvinnik, D.~O'Kane, P.~Hobson, S.~Lukauskas, D.~C. Gemperline,
  T.~Augspurger, Y.~Halchenko, J.~B. Cole, J.~Warmenhoven, J.~de~Ruiter,
  C.~Pye, S.~Hoyer, J.~Vanderplas, S.~Villalba, G.~Kunter, E.~Quintero,
  P.~Bachant, M.~Martin, K.~Meyer, A.~Miles, Y.~Ram, T.~Yarkoni, M.~L.
  Williams, C.~Evans, C.~Fitzgerald, Brian, C.~Fonnesbeck, A.~Lee, A.~Qalieh,
  mwaskom/seaborn: v0.8.1 (Sep. 2017).
\newblock \href {https://doi.org/10.5281/zenodo.883859}
  {\path{doi:10.5281/zenodo.883859}}.

\bibitem{plotly}
{Plotly Technologies Inc.}, \href{https://plot.ly}{Collaborative data science},
  Plotly Technologies Inc., Montreal, QC, 2015.
\newline\urlprefix\url{https://plot.ly}

\bibitem{gates2019clusim}
A.~J. Gates, Y.-Y. Ahn, Clusim: a python package for calculating clustering
  similarity, J. Open Source Softw. 4~(35) (2019) 1264.
\newblock \href {https://doi.org/https://doi.org/10.21105/joss.01264}
  {\path{doi:https://doi.org/10.21105/joss.01264}}.

\bibitem{kim2005systematic}
D.-H. Kim, H.~Jeong, Systematic analysis of group identification in stock
  markets, Phys. Rev. E 72~(4) (2005) 046133.
\newblock \href {https://doi.org/https://doi.org/10.1103/PhysRevE.72.046133}
  {\path{doi:https://doi.org/10.1103/PhysRevE.72.046133}}.

\bibitem{manning2008introduction}
C.~D. Manning, P.~Raghavan, H.~Sch{\"u}tze, Introduction to information
  retrieval, Cambridge Univ. Press, New York, 2008.

\bibitem{wu2009adapting}
J.~Wu, H.~Xiong, J.~Chen, Adapting the right measures for $k$-means clustering,
  in: Proceedings of the 15th ACM SIGKDD international conference on Knowledge
  discovery and data mining, 2009, pp. 877--886.
\newblock \href {https://doi.org/https://doi.org/10.1145/1557019.1557115}
  {\path{doi:https://doi.org/10.1145/1557019.1557115}}.

\bibitem{romano2016adjusting}
S.~Romano, N.~X. Vinh, J.~Bailey, K.~Verspoor,
  \href{http://jmlr.org/papers/v17/15-627.html}{Adjusting for chance clustering
  comparison measures}, J. Mach. Learn. Res. 17~(1) (2016) 4635--4666.
\newline\urlprefix\url{http://jmlr.org/papers/v17/15-627.html}

\bibitem{riker1982two}
W.~H. Riker, The two-party system and {Duverger's} law: An essay on the history
  of political science, Am. Political Sci. Rev. 76~(4) (1982) 753--766.

\end{thebibliography}

\end{document}